\documentclass[12pt]{article}
\usepackage{amsmath}
\usepackage{cite}

\csname @addtoreset\endcsname{equation}{section}
\newcommand{\eq}{\begin{equation}}
\newcommand{\feq}{\end{equation}}
\newcommand{\eqn}{\begin{eqnarray}}
\newcommand{\feqn}{\end{eqnarray}}
\newcommand{\arr}{\begin{eqnarray*}}
\newcommand{\farr}{\end{eqnarray*}}

\newcommand{\fft}[2]{{\frac{#1}{#2}}}
\newcommand{\ft}[2]{{\textstyle\frac{#1}{#2}}}

\begin{document}

\begin{titlepage}
\begin{flushright}
CAMS/04-01\\
MCTP-04-27\\
hep-th/0405171
\end{flushright}

\vspace{.3cm}
\begin{center}

\renewcommand{\thefootnote}{\fnsymbol{footnote}}
{\Large \bf Mass in anti-de Sitter spaces}\\

\vspace{1cm}

{\large \bf {James T. Liu$^1$\footnote{email: jimliu@umich.edu} and
W.~A.~Sabra$^2$\footnote{email: ws00@aub.edu.lb}}}\\

\renewcommand{\thefootnote}{\arabic{footnote}}
\setcounter{footnote}{0}

\vspace{1cm}

{\small
$^1$ {Michigan Center for Theoretical Physics,\\
Randall Laboratory of Physics, The University of Michigan,\\
Ann Arbor, MI 48109--1120}\\
\vspace*{0.4cm}
$^2$ Center for Advanced Mathematical Sciences (CAMS)\\
and\\
Physics Department, American University of Beirut, Lebanon.}
\end{center}

\vspace{1.5cm}
\begin{center}
{\bf Abstract}
\end{center}

The boundary stress tensor approach has proven extremely useful in
defining mass and angular momentum in asymptotically anti-de Sitter
spaces with CFT duals.  An integral part of this method is the use of
boundary counterterms to regulate the gravitational action and stress
tensor.  In addition to the standard gravitational counterterms, in the
presence of matter we advocate the use of a finite counterterm
proportional to $\phi^2$ (in five dimensions).  We demonstrate that this
finite shift is necessary to properly reproduce the expected mass/charge
relation for $R$-charged black holes in AdS$_5$.

\end{titlepage}


\section{Introduction}

\label{sec:intro}

While the notion of mass is perhaps intuitively obvious, much of this 
intuition is related to flat space, where mass may be used to label
representations of the Poincar\'e group.  Once we consider curved space,
some of this intuition falls apart.  Of course, the idea of mass as a
source of curvature is an essential component of general relativity.
Nevertheless, in the absence of Poincar\'e symmetry, mass can no longer
be defined in a straightforward manner.

In fact, in a closed spacetime, there can be no intrinsic meaning to the
mass of the universe in much the same way as there cannot be any net charge
in a closed space.  On the other hand, there has been a long history of
defining mass for spaces with an asymptotic region.  Perhaps one of the
best known prescriptions is that of Arnowitt, Deser and Misner (ADM), which
may be most straightforwardly applied in asymptotically flat spacetimes.  This
is essentially equivalent to reading off the mass from the Newtonian potential,
$\Phi(r)\sim -M/r$, where $\Phi(r)$ may may be extracted from the
time-time component of the metric, $g_{tt}\sim-(1+2\Phi(r))$.

In general, the ADM prescription can also be applied to spacetimes with
non-flat asymptotic regions, such as asymptotically anti-de Sitter (AdS)
spaces \cite{Abbott:1981ff}.  In such cases, the mass may be extracted by
comparison to a reference
({\it e.g.}~vacuum AdS) background.  However, care must be taken to ensure
that the deviation from the reference background is sufficiently well
controlled.  This task is often made difficult in practice because one
must control the reparametrization invariance of both deformed and undeformed
backgrounds to ensure a well defined result.  A similar approach to mass
has been taken by Brown and York \cite{Brown:1993br} in defining a quasilocal
stress tensor through the variation of the gravitational action
\begin{equation}
T^{ab}=\fft2{\sqrt{-h}}\fft{\delta S}{\delta h_{ab}},
\end{equation}
where $h_{ab}$ is the boundary metric.  In general, $T^{ab}$ diverges as
the boundary is pushed off to infinity, and hence a background subtraction
is again necessary.

More recently, an alternative procedure has been demonstrated where the
boundary stress tensor may be regulated by the introduction of appropriate
boundary counterterms \cite{Balasubramanian:1999re}.  The advantage of this
method is that the regulated gravitational action and resulting boundary
stress tensor may be obtained directly for the background at hand, without
having to introduce a somewhat artificial reference background.  This
counterterm method has become quite standard when applied to AdS/CFT, as the
boundary counterterms have a natural interpretation as conventional field
theory counterterms that show up in the dual CFT.

In general, it is only necessary to introduce a handful of boundary
counterterms in order to cancel divergences in the gravitational action.
For example, in AdS$_5$, only two counterterms are necessary.  However,
one could equally well add in an arbitrary amount of {\it finite}
counterterms.  While this would certainly change the values of the action
integral and corresponding boundary stress tensor, this has a natural
interpretation in the dual CFT as simply the usual freedom to change
renormalization prescriptions.

Although one is in principle free to choose any desired prescription, some
are perhaps better motivated than others.  For example, in a gauge theory,
one tends to avoid non-gauge invariant regulators, and in supersymmetric
theories, one generally chooses a `supersymmetric' scheme.  While the
introduction of finite counterterms has often been overlooked, this can
lead to somewhat surprising results.  In particular, it was shown in
\cite{Buchel:2003re} that, in the absence of finite counterterms, single
$R$-charged black holes in AdS$_5$ obey a non-linear mass/charge relation,
$M\sim \fft32\mu+q - \fft13g^2q^2$, where $\mu$ is the non-extremality
parameter and $g=1/\ell$ is the inverse AdS radius.  While nothing prevents
us from taking this as a definition of mass, it nevertheless appears to be
in conflict with the BPS expectation that $M\ge|q|$.

In this paper, we propose to include a finite counterterm related to the
scalar fields, and in doing so will recover the expected linear relation
$M\sim\fft32\mu+q$.  We also demonstrate that for three-charge AdS$_5$
black holes in the STU model, the mass/charge relation remains linear,
namely $M\sim\fft32\mu+q_1+q_2+q_3$.  The boundary stress tensor method
can also be applied to the newly constructed Gutowski-Reall black holes
in AdS$_5$ \cite{gutreal0,gutreal}.  We compute the masses of these solutions
and demonstrate equivalence with the results obtained in \cite{gutreal}
using the Ashtekar and Das approach \cite{ashdas}.

We begin in section 2 with a review of the boundary counterterm procedure.
While this is by now familiar, we find it useful here to set the notation
and prepare the groundwork for the subsequent calculations.  In section 3
we include matter fields (general scalars and vectors) and in section 4
we complete the regulation procedure by introducing a finite $\phi^2$
counterterm.  We verify in section 5 that this counterterm results in
the linear mass relation mentioned above.  Finally, we examine the
Gutowski-Reall black holes in section 6 and conclude in section 7.


\section{The stress tensor for pure gravity}

Before considering the matter coupled system, we briefly review the boundary
counterterm method for a purely gravitational theory
\cite{Balasubramanian:1999re}.  We work in five dimensions with a negative
cosmological constant, so that the Einstein action may be written as
\begin{equation}
S[g_{\mu \nu }]=S_{\rm bulk}+S_{\rm GH}
=-\frac{1}{16\pi G_{5}}\int_{\mathcal{M}}d^{5}x\sqrt{-g}\left[
R+12g^{2}\right]
+\frac{1}{8\pi G_{5}}\int_{\partial \mathcal{M}}d^{4}x\sqrt{-h}~\Theta,
\label{eq:d5act}
\end{equation}
where $g$ is the inverse radius of AdS$_5$.  The Gibbons-Hawking surface
term is included to ensure a proper variational principle for a spacetime
$\mathcal{M}$ with boundary $\partial \mathcal{M}$. Here, $\Theta$ is the
trace of the extrinsic curvature $\Theta ^{\mu\nu}$ of the boundary, defined
by 
\begin{equation}
\Theta^{\mu\nu}=-\frac{1}{2}\left(\nabla^{\mu}n^{\nu}+\nabla^{\nu}n^{\mu}
\right),
\end{equation}
where $n^{\mu}$ is the outward-pointing normal on $\partial \mathcal{M}$.

In the holographic context, it is natural to single out a radial coordinate
$r$, and thus we decompose the bulk five-dimensional metric according to
\begin{equation}
ds_5^{2}=N^{2}dr^{2}
+h_{ab}\bigl(dx^{a}+V^{a}dr\bigr)\bigl(dx^{b}+V^{b}dr\bigr).
\end{equation}
This is essentially an ADM decomposition, except that here the radial
coordinate $r$ plays the r\^ole of time.  Furthermore we will choose $r$
so that the boundary,
$\partial\mathcal{M}$, is reached as $r\to\infty$.  The four-dimensional
metric $h_{ab}$ then represents the induced metric on $\partial\mathcal{M}$.
Following \cite{Brown:1993br}, the quasi-local stress tensor on the surface
$\partial \mathcal{M}$ is then defined through the variation of the
gravitational action with respect to the boundary metric $h_{ab}$
\begin{equation}
T^{ab}=\frac{2}{\sqrt{-h}}\frac{\delta S}{\delta h_{ab}}=\frac{1}{8\pi G_{5}}
\left(\Theta ^{ab}-\Theta h^{ab}\right).
\label{eq:tab}
\end{equation}
Given $T^{ab}$, it is possible to extract the ADM mass and momentum as
appropriate conserved quantities.  To do so, we foliate the boundary
spacetime $\partial \mathcal{M}$ by spacelike surfaces $\Sigma $ with
metric $\sigma _{\alpha\beta}$, so that 
\begin{equation}
ds_4^2\equiv h_{ab}dx^{a}dx^{b}=-N_{\Sigma }^{2}dt^{2}+\sigma_{\alpha\beta}
\bigl(dx^{\alpha}+V_{\Sigma}^{\alpha}dt\bigr)
\bigl(dx^{\beta}+V_{\Sigma}^{\beta}dt\bigr).
\label{eq:bdymet}
\end{equation}
The conserved charges are then obtained by integrating the time component
of the conserved stress tensor over the three-dimensional surface $\Sigma$.
More precisely, for an isometry of the boundary geometry generated by
a Killing vector $\xi^a$, the corresponding conserved charge is given by
\begin{equation}
Q_{\xi }=\int_{\Sigma}dx^{3}\sqrt{\sigma }\left( u^{a}T_{ab}\xi ^{b}\right),
\label{eq:consq}
\end{equation}
where $u^{a}$ is the timelike unit normal to the surface $\Sigma$.  For a
time-translationally invariant spacetime, we take the Killing vector to be
$\xi ^{a}=N_{\Sigma }u^{a}$, in which case the conserved charge $Q_\xi$
corresponds to the total energy of the spacetime.

In general, it can be shown that the stress tensor defined in this matter
(as well as the on-shell value of the action) diverges when the surface
$\partial \mathcal{M}$ is pushed to infinity.  While \cite{Brown:1993br}
removes this divergence through background subtraction, the method of
\cite{Balasubramanian:1999re} is to instead regulate the action,
(\ref{eq:d5act}), through the addition of boundary counterterms,
$S_{\rm ct}[h_{ab}]$.  This also yields a counterterm addition to the
stress tensor,
\begin{equation}
T_{\rm reg}^{ab}=\fft1{8\pi G_5}(\Theta^{ab}-\Theta h^{ab})+\fft2{\sqrt{-h}}
\fft{\delta S_{ct}}{\delta h_{ab}}.
\label{eq:treg}
\end{equation}
Only two counterterms, of the forms
\begin{equation}
S_{1} =\frac{1}{8\pi G_{5}}\int d^{4}x\sqrt{-h},\qquad
S_{2} =\frac{1}{8\pi G_{5}}\int d^{4}x\sqrt{-h}\mathcal{R},
\label{eq:gct}
\end{equation}
are necessary for regulating the divergences of the gravitational action,
(\ref{eq:d5act}).  Here ${\cal R}$ is the scalar curvature of the boundary
metric (\ref{eq:bdymet}).  The resulting action has the form
\begin{equation}
S_{\rm reg}=S_{\rm bulk}+S_{\rm GH}+3gS_1+(4g)^{-1}S_2.
\label{eq:regact}
\end{equation}
In addition, the counterterms contribute
\begin{equation}
T_1^{ab}=\fft1{8\pi G_5}h^{ab},\qquad
T_2^{ab}=\fft1{8\pi G_5}(2{\cal R}^{ab}-{\cal R}h^{ab}),
\label{eq:t1t2}
\end{equation}
to the regulated stress tensor.

\subsection{Mass of the Schwarzschild-AdS spacetimes}

To illustrate the above general discussion, we review the case of
Schwarzschild-AdS$_5$, which has attracted much previous attention
as the spacetime corresponding to non-extremal D3-branes.  In five
dimensions, the metric may be written as 
\begin{equation}
ds^{2}=-f(r)dt^{2}+\frac{dr^{2}}{f(r)}+r^{2}d\Omega _{3}^{2},
\end{equation}
where $f(r)=1-(r_{0}/r)^{2}+g^{2}r^{2}$.  We implicitly define $r_+$ to
be the location of the horizon, given by $f(r_+)=0$.

When evaluated on shell, the bulk action may be re-expressed in terms of a
surface integral.  We find 
\begin{equation}
I_{\mathrm{bulk}}=\frac{\beta \omega _{3}}{8\pi G_{5}}\left(
r^{2}(f-1)+r_{+}^{2}\right),
\label{eq:ibulk0}
\end{equation}
where we use $I$ to denote the value of the Euclidean action integral. Here,
$\beta=2\pi/T$ is the periodicity along the time direction and
$\omega_{3}=2\pi^2$ is the volume of the unit $3$-sphere.  Note that, in
the absence of matter, the action integral (\ref{eq:ibulk0}) is easily
obtained through the substitution of the trace of the Einstein equation,
$R=-20g^{2}$, into $S_{\mathrm{bulk}}$ to obtain
\begin{equation}
I_{\rm bulk}=-{\frac{1}{16\pi G_{5}}}\int d^{5}x\sqrt{-g}[-8g^{2}]
={\frac{\beta \omega _{3}}{8\pi G_{5}}}g^{2}(r^{4}-r_{+}^{(4)}),
\end{equation}
which is equivalent to (\ref{eq:ibulk0}) when the expression for $f(r)$ is
taken into account. However, as shown in the following section, the expression
(\ref{eq:ibulk0}) is more general, and continues to hold when matter is
added to the system.

In addition to $I_{\rm bulk}$, the Gibbons-Hawking boundary term gives
the contribution 
\begin{equation}
I_{\mathrm{GH}}=-\frac{\beta \omega _{3}}{8\pi G_{5}}\bigl({\textstyle\frac{1%
}{2}}r^{3}f^{\prime }+3r^{2}f\bigr).
\end{equation}
Thus the complete action is given by 
\begin{equation}
I_{\mathrm{GH}}+I_{\mathrm{bulk}}=\frac{\beta \omega _{3}}{8\pi G_{5}}\bigl(%
-3r^{2}+r_{+}^{2}-3g^{2}r^{4}+{\textstyle}r_{0}^{2}\bigr),
\end{equation}
where we have substituted in the explicit form of $f$.

While the on-shell action diverges like $r^4$ as we approach the boundary
$r\to\infty$, this divergence is removed by the addition of the counterterms
(\ref{eq:gct}).  The appropriately regulated action, (\ref{eq:regact}), is
given by \cite{Emparan:1999pm,Balasubramanian:1999re} 
\begin{equation}
I_{\rm reg}=
\frac{\beta\pi}{4G_{5}}\left(r_{+}^{2}-\frac{1}{2}r_{0}^{2}+\frac{3}{8g^{2}}
\right),
\label{eq:reg5i}
\end{equation}
and remains finite.  Likewise, the counterterms also lead to a finite stress
tensor.  Using (\ref{eq:treg}) and (\ref{eq:t1t2}), one finds
\begin{equation}
T_{tt}=\frac{1}{8\pi G_5r^{2}}\left(\frac{3}{8g}+\frac{3gr_{0}^{2}}{2}\right),
\end{equation}
resulting the familiar Schwarzschild-AdS$_{5}$ energy
\cite{Balasubramanian:1999re} 
\begin{equation}
E=\frac{3\pi }{8G_{5}}\left( r_{0}^{2}+\frac{1}{4g^{2}}\right),
\label{eq:bsmass}
\end{equation}
which naturally includes the CFT Casimir energy in addition to the
non-extremality parameter $r_0$.


\section{Addition of the matter sector}

In order to examine the mass of charged black hole solutions, we must first
extend the standard counterterm procedure by introducing a matter sector to
the bulk action. Since the action is no longer that of pure gravity, it may
now be necessary to include additional local counterterms on $\partial
\mathcal{M}$ constructed out of the boundary values of the matter fields in
order to cancel all divergences.

Although we eventually turn to solutions of gauged ${\cal N}=2$ supergravity
in five dimensions, we first consider the a general matter coupled gravity
system with action 
\begin{eqnarray}
S[g_{\mu \nu },\phi ^{i},A_{\mu }^{I}]\!&=&\!-{\frac{1}{16\pi G_{5}}}
\int_{\mathcal{M}}d^{5}x\sqrt{-g}[R-{\textstyle\frac{1}{2}}g(\phi
)_{ij}\partial _{\mu }\phi ^{i}\partial ^{\mu }\phi ^{j}-{\textstyle\frac{1}{%
4}}G_{IJ}(\phi )F_{\mu \nu }^{I}F^{\mu \nu \,J}-V(\phi )]  \notag \\
&&+\frac{1}{8\pi G_{5}}\int_{\partial \mathcal{M}}d^{4}x\sqrt{-h}~\Theta .
\label{eq:bulka}
\end{eqnarray}
To evaluate the on-shell value of the bulk action, we note that the Einstein
equation, written in Ricci form, is given by 
\begin{equation}
R_{\mu \nu }=\ft12g(\phi )_{ij}\partial _{\mu }\phi^{i}\partial_{\nu}\phi^{j}
+\ft12G_{IJ}(F_{\mu\lambda}^{I}F_{\nu}{}^{\lambda\,J}-\ft16g_{\mu\nu}
F_{\rho\sigma}^{I}F^{\rho\sigma\,J})+\ft13g_{\mu\nu}V.
\label{eq:eins}
\end{equation}
Taking the trace of this equation to obtain $R$, and substituting it into
the action integral gives 
\begin{equation}
I_{\mathrm{bulk}}=-{\frac{1}{16\pi G_5}}\int d^{5}x\sqrt{-g}[-\ft16
G_{IJ}F_{\mu \nu }^{I}F^{\mu \nu \,J}+\ft23V].
\label{eq:ibgen}
\end{equation}

While this is a simplification of the action integral, it appears to be as
far as we may proceed without further input. Thus we now focus on static
electrically charged black hole solutions, and take an ansatz of the form 
\begin{eqnarray}
&&ds^{2}=-e^{-4B(r)}f(r)dt^{2}+e^{2B(r)}\left( {\frac{dr^{2}}{f(r)}}%
+r^{2}d\Omega _{3}^{2}\right), \notag\\
&&\phi ^{i}=\phi ^{i}(r),\qquad A_{t}^{I}=A_{t}^{I}(r),
\label{eq:bhans}
\end{eqnarray}
where the $3$-sphere may be parametrized as 
\begin{equation}
d\Omega_3^{2}=d\psi ^{2}+\sin ^{2}\psi d\Omega_{2}^{2}.
\end{equation}
In this case, the $R_{\psi\psi}$ component of the Einstein equation,
(\ref{eq:eins}), yields 
\begin{equation}
2R_{\psi }^{\psi }=-{\textstyle\frac{1}{6}}G_{IJ}F_{\mu \nu }^{I}F^{\mu \nu
\,J}+{\textstyle\frac{2}{3}}V,
\end{equation}
which has the same form as the integrand of (\ref{eq:ibgen}). This gives a
simple result for the action integral
\begin{equation}
I_{\mathrm{bulk}}=-{\frac{1}{8\pi G_{5}}}\int d^{5}x\sqrt{-g}R_{\psi}^{\psi},
\label{eq:simpa}
\end{equation}
provided we follow the ansatz (\ref{eq:bhans}). Working out the $R_{\psi\psi}$
component explicitly, we obtain 
\begin{eqnarray}
I_{\mathrm{bulk}}&=&{\frac{\beta\omega_{3}}{8\pi G_{5}}}\int dr{\frac{d}{dr}}
\bigl[r^{3}fB^{\prime }+r^{2}(f-1)\bigr]  \notag \\
&=&\frac{\beta \omega _{3}}{8\pi G_{5}}\left( r^{3}fB^{\prime}
+r^{2}(f-1)+r_{+}^{2}\right) ,
\end{eqnarray}
where in the last line we have taken the range of $r$ to be from the horizon 
$r_{+}$ to the finite but large value $r$ where we cut off the space.

To evaluate the Gibbons-Hawking surface term, we start with the unit normal
in the $r$ direction, $n^{r}=e^{-B}f^{{\frac{1}{2}}}$. Evaluating its
divergence yields 
\begin{equation}
\Theta =-\nabla _{\mu }n^{\mu }=-e^{-B}f^{{\frac{1}{2}}}\left( B^{\prime }
+{\frac{f^{\prime}}{2f}}+{\frac{3}{r}}\right) ,
\end{equation}
so that 
\begin{equation}
I_{\mathrm{GH}}=-\frac{\beta \omega _{3}}{8\pi G_{5}}\left( r^{3}fB^{\prime}
+\ft12r^{3}f^{\prime}+3r^{2}f\right).
\end{equation}
Curiously, the $B^{\prime}$ dependent terms in $I_{\mathrm{bulk}}$ and
$I_{\mathrm{GH}}$ cancel when added together. We find 
\begin{equation}
I_{GH}+I_{bulk}=\frac{\beta\omega_{3}}{8\pi G_{5}}
\left(-2r^{2}f-\ft12r^{3}f^{\prime}-r^{2}+r_{+}^{2}\right) .
\label{eq:unreg}
\end{equation}
This expression as it stands is divergent, and must be regulated by an
appropriate counterterm subtraction. However, we emphasize that this
expression includes all effects of the scalars and gauge fields of (\ref
{eq:bulka}), although they do not show up explicitly here. It is remarkable
that the unregulated action only depends explicitly on the `blackening
function' $f(r)$ in (\ref{eq:bhans}). However, as a solution to the Einstein
equation, $f$ naturally includes residual information of all appropriate
scalar and gauge charges.

For the metric ansatz, (\ref{eq:bhans}), the boundary counterterms $S_1$
and $S_2$ take the simple form
\begin{equation}
I_{1}={\frac{\beta \omega _{3}}{8\pi G_{5}}}r^{3}f^{{\frac{1}{2}}}e^{B}, 
\qquad
I_{2}={\frac{3\beta \omega _{3}}{4\pi G_{5}}}rf^{{\frac{1}{2}}}e^{-B},
\end{equation}
so that the regulated action integral, (\ref{eq:regact}), is given by
\begin{equation}
I_{\rm reg}=\frac{\beta\omega_{3}}{8\pi G_{5}}
\left(-2r^{2}f-\ft12r^{3}f^{\prime}-r^{2}+r_{+}^{2}
+3gr^3f^{\fft12}e^B+\ft32g^{-1}rf^{\fft12}e^{-B}\right).
\label{eq:i5bh}
\end{equation}
Although this is not manifestly finite, we demonstrate explicitly
that it is indeed so for $R$-charged black holes in AdS$_5$.

\subsection{$R$-charged black holes in AdS$_5$}

We now turn to the examination of $R$-charged black holes.
These electrically charged black holes are static stationary solutions
to gauged ${\cal N}=2$ supergravity, and have a metric of the form 
\cite{Behrndt:1998jd}
\begin{equation}
ds^{2}=-\mathcal{H}^{-2/3}fdt^{2}+\mathcal{H}^{1/3}\left( \frac{dr^{2}}{f}%
+r^{2}d\Omega _{3}^{2}\right) ,
\label{eq:rbhmet}
\end{equation}
where 
\begin{equation}
f=1-{\frac{r_{0}^{2}}{r^{2}}}+g^{2}r^{2}\mathcal{H}.
\label{eq:fdef}
\end{equation}
The `harmonic function' $\mathcal{H}$ is related to $e^{2B}$ of the ansatz
(\ref{eq:bhans}) by $\mathcal{H}=e^{6B}$. In the STU model, $\mathcal{H}$ is
given by the product of three harmonic functions 
\begin{equation}
\mathcal{H}=H_{1}H_{2}H_{3}=\left(1+{\frac{q_{1}}{r^{2}}}\right)
\left(1+{\frac{q_{2}}{r^{2}}}\right)\left(1+{\frac{q_{3}}{r^{2}}}\right).
\label{eq:calhdef}
\end{equation}
However, in general, we still expect $\mathcal{H}$ to have a large $r$
expansion of the form 
\begin{equation}
\mathcal{H}=1+{\frac{Q^{(1)}}{r^{2}}}+{\frac{Q^{(2)}}{r^{4}}}+
{\frac{Q^{(3)}}{r^{6}}}+\cdots .
\label{eq:Qidef}
\end{equation}

It is now straightforward to substitute the metric functions $f(r)$
and ${\cal H}(r)$ into the regulated action integral, (\ref{eq:i5bh}).
Up to terms that vanish in the limit $r\to\infty$, we obtain the finite
expression
\begin{equation}
I_{\rm reg}=\frac{\beta\pi}{4G_{5}}
\left(r_{+}^{2}-{\frac{1}{2}}r_{0}^{2}+\frac{3}{8g^{2}}
-g^{2}\left(\frac{Q^{(1)\,2}}{3}-Q^{(2)}\right) \right).
\label{eq:reg5q}
\end{equation}
This expression is the generalization of (\ref{eq:reg5i}) to the case of
$R$-charged black holes, where the charges are given by $Q^{(1)}=\sum_iq_i$
and $Q^{(2)}=\sum_{i<j} q_iq_j$. Note that this expression is obtained
directly from the metric (\ref{eq:rbhmet}), without even specifying the
gauge fields and scalars associated with the solution.

As can be seen from (\ref{eq:reg5q}), the black hole charges enter
non-linearly in the action integral.  In particular, for the single
charged black hole ($Q^{(2)}=0$) this expression reduces to that derived
previously in \cite{Buchel:2003re}.  Although there is nothing inherently
wrong with the nonlinear charge behavior, it is somewhat unexpected,
especially considering that it remains nonlinear in the BPS limit.  Of
course, these black holes actually become singular in the limit.  But
nevertheless, the formal BPS expression could have been expected to hold.
Indeed, it turns out that there is a simple means of removing the
nonlinearity in (\ref{eq:reg5q}) through the introduction of finite boundary
counterterms.  This is what we now proceed to demonstrate.


\section{Addition of finite counterterms}

In general, boundary counterterms have been introduced as a means of
regulating divergences in the gravitational action. However, we wish to
emphasize here that nothing prevents us from introducing \textit{finite}
counterterms as well. Such expressions yield a finite renormalization of the
gravitational action, and are hence dual to finite shifts in the
renormalization of the CFT. As a result, they are may be viewed as
generating shifts between different renormalization prescriptions of the CFT.

Since we have introduced the matter fields $\phi^i$ and $A_\mu^I$ into the
action (\ref{eq:bulka}), it is natural to construct local boundary
counterterms such as 
\begin{eqnarray}
&&S_{\phi^2}=\frac{1}{8\pi G_{5}}\int_{\partial\mathcal{M}}d^{4}x\sqrt{-h}
\,g_{ij}\phi^i\phi^j,\qquad
S_{\partial\phi^2}=\frac{1}{8\pi G_{5}}
\int_{\partial\mathcal{M}}d^{4}x \sqrt{-h}\,g_{ij}\partial_a\phi^i
\partial^a\phi^j,  \notag \\
&&S_{F^2}={\frac{1}{8\pi G_5}}\int_{\partial\mathcal{M}}d^4x\sqrt{-h}\,
G_{IJ}F_{ab}^IF^{ab\,J},
\label{eq:ftct}
\end{eqnarray}
where the $a$, $b$ indices correspond to the boundary surface $\partial 
\mathcal{M}$. In particular, radial $\partial_r$ derivatives are absent, as
they are not local to the boundary.

For the spherically symmetric black holes, the fields are only functions of
the radial coordinate $r$.  Hence the two-derivative counterterms in
(\ref{eq:ftct}) will not contribute. As a result, we consider only
$I_{\phi ^{2}}$, which takes the form 
\begin{equation}
I_{\phi^{2}}=\frac{\beta\omega_3}{8\pi G_{5}}r^{3}f^{\frac{1}{2}}e^{B}
(g_{ij}\phi^{i}\phi^{j}).
\label{eq:iphi2}
\end{equation}
So far, the analysis has been completely general, at least for this class
of spherically symmetric and stationary solutions.  However, at this stage,
it is necessary to provide the explicit asymptotic form of the scalars
corresponding to the black hole metric of (\ref{eq:rbhmet}).

To proceed, we consider the specific example of the STU model. Here, there
are three $U(1)$ gauge fields and two scalars, with the scalars defined by 
\begin{eqnarray}
X^{1} &=&e^{-\frac{1}{\sqrt{6}}\phi _{1}-\frac{1}{\sqrt{2}}\phi
_{2}}=H_{1}^{-1}\mathcal{H}^{\frac{1}{3}},  \notag \\
X^{2} &=&e^{-\frac{1}{\sqrt{6}}\phi _{1}+\frac{1}{\sqrt{2}}\phi
_{2}}=H_{2}^{-1}\mathcal{H}^{\frac{1}{3}},  \notag \\
X^{3} &=&e^{\frac{2}{\sqrt{6}}\phi _{1}}=H_{3}^{-1}\mathcal{H}^{\frac{1}{3}},
\end{eqnarray}
where $\mathcal{H}=H_{1}H_{2}H_{3}$, so that $X^{1}X^{2}X^{3}=1$. The two
independent scalars $\phi _{1}$ and $\phi _{2}$ may be re-expressed as 
\begin{eqnarray}
\phi _{1} &=&\frac{1}{\sqrt{6}}\left( \log H_{1}+\log H_{2}-2\log
H_{3}\right) , \\
\phi _{2} &=&\frac{1}{\sqrt{2}}\left( \log H_{1}-\log H_{2}\right) .
\end{eqnarray}
This gives in turn 
\begin{equation}
\vec{\phi}\,^{2}=\phi_{1}^{2}+\phi_{2}^{2}=\frac{1}{r^{4}}
\left(\frac{2}{3}Q^{(1)\,2}-2Q^{(2)}\right)+\cdots .
\label{eq:phisq}
\end{equation}
Substituting this into (\ref{eq:iphi2}) yields a finite contribution 
\begin{equation}
{\frac{1}{g}}I_{\phi^{2}}=\frac{\beta\pi}{4G_{5}}
\left(\frac{2}{3}Q^{(1)\,2}-2Q^{(2)}\right) ,
\end{equation}
which has the exact same charge dependence as the finite part of the action,
(\ref{eq:reg5q}). As a result, it may be used as a finite counterterm to
completely cancel the charge dependence of the action.  The regulated action
integral, including finite counterterm
\begin{equation}
I_{\rm reg}+\frac{g}{2}I_{\phi^{2}}
=\frac{\beta\pi}{4G_{5}}\left(r_{+}^{2}-{\frac{1}{2}}
r_{0}^{2}+\frac{3}{8g^{2}}\right),
\label{eq:reg5full}
\end{equation}
is then identical to that of the Schwarzschild-AdS solution,
(\ref{eq:reg5i}). In fact, we are advocating the use of the full
counterterm action
\begin{equation}
I_{\rm complete} = I_{\rm bulk}+I_{\rm GH}+3gI_1+\fft1{4g}I_2
+\fft{g}2I_{\phi^2},
\label{eq:fcta}
\end{equation}
for black holes in AdS$_5$ with or without $R$ charge. For the latter, of
course, the $I_{\phi^{2}}$ counterterm vanishes. However, we may view
this counterterm action as universal, with all coefficients independent
of charge.


\section{The regulated boundary stress tensor}

In the previous section, we have shown that an appropriate counterterm
prescription exists for five-dimensional $R$-charged black holes that
preserves the standard expression (\ref{eq:reg5i}) for the action integral
independent of charge. We now turn to the calculation of the boundary
stress tensor and the extraction of the ADM energy.

We begin with the unregulated stress tensor, (\ref{eq:tab}), given by 
\begin{equation}
T^{ab}=\frac{1}{8\pi G_{5}}\left(\Theta^{ab}-\Theta h^{ab}\right) .
\label{eq:bst}
\end{equation}
For the metric (\ref{eq:bhans}), the extrinsic curvature takes the form 
\begin{eqnarray}
\Theta ^{tt} &=&-\left(-2B^{\prime }+\frac{f^{\prime }}{2f}\right)
h^{tt}e^{-B}f^{\frac{1}{2}},  \notag \\
\Theta ^{\alpha \beta } &=&-\left( B^{\prime }+\frac{1}{r}\right)
h^{\alpha\beta}e^{-B}f^{\frac{1}{2}},
\end{eqnarray}
so that 
\begin{equation}
\Theta =-\left( B^{\prime }+\frac{3}{r}+\frac{f^{\prime }}{2f}\right)
e^{-B}f^{\frac{1}{2}}.
\end{equation}
Substituting these expressions into (\ref{eq:bst}) gives 
\begin{equation}
T_{tt}=\frac{g}{8\pi G_{5}}\left(-3g^{2}r^{2}-Q^{(1)}g^{2}-\frac{9}{2}+
\frac{1}{r^{2}}\left(\frac{9r_{0}^{2}}{2}+3Q^{(1)}-\frac{9}{8g^{2}}\right)
\right) .
\end{equation}
At the same time, the local gravitational counterterms, $S_1$ and $S_2$,
give rise to the contribution
\begin{eqnarray}
\tilde{T}_{tt} &=&-\frac{g}{8\pi G_5}3g_{tt}\left( 1+\frac{e^{-2B}}
{2g^{2}r^{2}}\right)\\
&=&\frac{g}{8\pi G_5}\left(3g^{2}r^{2}+Q^{(1)}g^{2}+\frac{9}{2}+
\frac{1}{r^{2}}\left(-2Q^{(1)}-3r_0^2+g^2\left(Q^{(2)}-\fft13Q^{(1)\,2}
\right)+\frac{3}{2g^2}\right)\right),\nonumber
\end{eqnarray}
so that the gravitationally regulated value of $T_{tt}$ is 
\begin{equation}
T_{tt}^{\rm reg}=\frac{g}{8\pi G_5r^{2}}\left(Q^{(1)}+\frac{3r_0^2}{2}
+g^{2}\left(Q^{(2)}-\frac{Q^{(1)\,2}}{3}\right)+\frac{3}{8g^2}\right) .
\end{equation}
While this expression yields a finite energy when inserted in
(\ref{eq:consq}), the term quadratic in charge gives rise to a
non-linear mass/charge relation, as first noted in \cite{Buchel:2003re}.
In fact, setting $Q^{(2)}=0$ reproduces the single-charge black hole
result of \cite{Buchel:2003re}.

Of course, the introduction of the finite counterterm $S_{\phi^2}$ also
shifts the stress tensor according to
\begin{equation}
T_{\phi^2}^{ab}=\fft1{8\pi G_5}h^{ab}(g_{ij}\phi^i\phi^j).
\end{equation}
For the STU model, the evaluation of $(g_{ij}\phi^i\phi^j)$ follows from 
(\ref{eq:phisq}).  Including $T_{\phi^2}^{ab}$ results in cancellation
of the nonlinear charge term, so that the fully regulated value of $T_{tt}$
takes on the simple form
\begin{equation}
T_{tt}^{\mathrm{complete}}=\frac{g}{8\pi G_5r^{2}}\left(Q^{(1)}
+\frac{3r_0^2}{2} +\frac{3}{8g^{2}}\right).
\end{equation}
Therefore the energy is given by 
\begin{eqnarray}
E &=&\frac{\pi}{4G_5}\left(Q^{(1)}+\frac{3r_0^2}{2}+\frac{3}{8g^{2}}\right) 
\notag \\
&=&\frac{3\pi}{8G_5}\left(r_0^2+\fft23q_{1}+\fft23q_{2}+\fft23q_{3}
+\frac{1}{4g^{2}}\right),
\label{eq:goodmass}
\end{eqnarray}
where in the last line we have explicitly written out the three charges of
the STU model. This energy generalizes the expression (\ref{eq:bsmass}) to the
case of charged black holes in AdS$_5$.

By adding a finite counterterm, $I_{\phi^2}$, we have been able to provide a
rigorous justification of the AdS$_5$ black hole mass originally given in 
\cite{Behrndt:1998jd}. We believe this expression is natural from the
viewpoint of thermodynamics, in that the three independent charges of the
STU model (or equivalently the three commuting $U(1)^3\subset SO(6)_R$
charges of the four-dimensional $\mathcal{N}=4$ theory) contribute linearly
to the mass in (\ref{eq:goodmass}). Note that the non-linear term $%
Q^{(2)}-Q^{(1)\,2}/3$ vanishes identically for the three equal charge black
hole. In this case, either mass expression yields the same result. In fact,
this must be true; since this black hole may be viewed as a solution of the
pure five-dimensional $\mathcal{N}=2$ supergravity, and there are no scalars
in this theory, the scalar counterterm cannot contribute to the mass.


\section{Gutowski-Reall Solutions}

Another example where the importance of the $S_{\phi^2}$ counterterm shows
up is in the case of the recently constructed Gutowski-Reall supersymmetric 
black hole solutions \cite{gutreal0,gutreal}.  Unlike the stationary
$R$-charged black holes investigated above which become singular in the
BPS limit, the Gutowski-Reall solutions maintain a regular horizon through
non-zero angular momentum.

The general rotating solution has a metric of the form \cite{gutreal}
\begin{equation}
ds^{2}=-f^{2}dt^{2}-2f^{2}w\,dt\,\sigma_{L}^{3}+f^{-1}g^{-1}dr^{2}
+\frac{r^{2}}{4}\left[ f^{-1}\left(\left(\sigma_{L}^{1}\right)^{2}
+\left(\sigma_{L}^{2}\right)^{2}\right)
+f^{2}h\left(\sigma_{L}^{3}\right)^{2}\right],
\label{eq:grmet}
\end{equation}
where the functions $f$, $g$, $w$ and $h$ are
\begin{eqnarray}
f &=&\left(1+\frac{\alpha_{1}}{r^{2}}+\frac{\alpha_{2}}{r^{4}}
+\frac{\alpha_{3}}{r^{6}}\right)^{-\frac{1}{3}},\qquad
g=\left(1+\frac{\alpha_{1}}{\ell^{2}}+\frac{r^{2}}{\ell^{2}}\right),\nonumber\\
w &=&-\frac{\epsilon r^{2}}{2\ell}\left(1+\frac{\alpha_{1}}{r^{2}}
+\frac{\alpha_{2}}{2r^{4}}\right),\qquad~~
h =f^{-3}g-\frac{4}{r^{2}}w^{2}.
\end{eqnarray}
Here, $\sigma_L^i$ are right-invariant 1-forms on SU(2) given by
\begin{eqnarray}
\sigma _{L}^{1} &=&\sin\phi\,d\theta-\cos\phi\sin\theta\,d\psi,\nonumber\\
\sigma _{L}^{2} &=&\cos\phi\,d\theta+\sin\phi\sin\theta\,d\psi,\nonumber\\
\sigma _{L}^{3} &=&d\phi+\cos\theta\,d\psi.
\end{eqnarray}
In addition, the scalars and gauge fields are given generally by
\begin{eqnarray}
X_{I} &=&f\left( \bar{X}_{I}+\frac{q_{I}}{r^{2}}\right),\nonumber\\
A^{I} &=&fX^{I}dt+\left(U^{I}+fwX^{I}\right)\sigma_{L}^{3},\nonumber\\
U^I&=&\fft{9\epsilon}{4\ell}C^{IJK}\bar{X}_J\left(\bar{X}_Kr^2+2q_K\right).
\end{eqnarray}
To avoid confusion over the gauge coupling $g$ versus the function $g(r)$,
we maintain the notation of \cite{gutreal} where $\ell$ denotes the AdS$_5$
radius.  These expressions simplify for the STU model, in which case
\begin{equation}
f=(H_1H_2H_3)^{-1/3},\qquad X_I=\ft13H_If,\qquad H_I=1+\fft{\mu_I}{r^2}.
\end{equation}
Note that
\begin{equation}
\alpha_1=\mu_1+\mu_2+\mu_3,\qquad
\alpha_2=\mu_1\mu_2+\mu_1\mu_3+\mu_2\mu_3,\qquad
\alpha_3=\mu_1\mu_2\mu_3,
\end{equation}
are analogous to the $Q^{(i)}$ of (\ref{eq:Qidef}).

Even in the STU model, where the scalar and gauge field behavior is
explicit, the analysis is somewhat complicated by rotation.  Foliating
the boundary metric according to (\ref{eq:bdymet}), we first rewrite
(\ref{eq:grmet}) as
\begin{equation}
ds^2=-\fft{g}{fh}dt^2+\fft{dr^2}{fg}+\fft{r^2}{4f}\left[
(\sigma_L^1)^2+(\sigma_L^2)^2+f^3h(\sigma_L^3-\fft4{r^2}\fft{w}h dt)^2
\right].
\end{equation}
This then allows us to introduce a natural vielbein basis
\begin{eqnarray}
&&e_0=g^{\fft12}(fh)^{-\fft12}dt,\qquad e_4=(fg)^{-\fft12}dr,\nonumber\\
&&e_1=\fft{r}2f^{-\fft12}\sigma_L^1,\qquad
e_2=\fft{r}2f^{-\fft12}\sigma_L^2,\qquad
e_3=\fft{r}2fh^{\fft12}(\sigma_L^3-\fft4{r^2}\fft{w}h dt).
\end{eqnarray}
Given this solution, we may compute the regulated action integral
(\ref{eq:fcta}) as well as the corresponding boundary stress tensor.

\subsection{The action integral}

While computation of the bulk action is in principle straightforward, the
simplification of (\ref{eq:simpa}) no longer follows due to the rotation.
Of course, one may still evaluate $I_{\rm bulk}$ directly from
(\ref{eq:ibgen}) and explicit knowledge of the solution.  Alternatively,
one can use the $R_{11}$ component of the Einstein equation to rewrite
(\ref{eq:simpa}) as
\begin{equation}
I_{\rm bulk}=-\fft1{8\pi G_5}\int d^5x\sqrt{-g}[R_{11}-\ft12G_{IJ}
F_{12}^IF_{12}^J],
\label{eq:rotiba}
\end{equation}
where we have also used the fact that the only non-vanishing vielbein
components of the field strength are $F_{04}^I$, $F_{12}^I$ and $F_{34}^I$.

For the STU model, the second term in (\ref{eq:rotiba}) has the form
\begin{equation}
G_{IJ}F_{12}^IF_{12}^J=\fft{f^4}{\ell^2r^8}(3\alpha_2^2-8\alpha_1\alpha_3).
\end{equation}
In addition, the Ricci component $R_{11}$ may be written as
\begin{equation}
R_{11}=\fft{f}{r^3}\left[\fft{d}{dr}\left(\fft12r^3 g\fft{f'}f+r^2(2-g)\right)
-2rhf^3\right].
\end{equation}
Combining these expressions and integrating from the horizon $r=0$ to a
large radial value $r$ yields
\begin{eqnarray}
I_{\rm bulk}&=&\fft{\beta\omega_3}{8\pi G_5}\Bigl[\fft{r^4}{\ell^2}
+\fft{2\alpha_1r^2}{3\ell^2}+\fft{\alpha_1}3+\fft{2\alpha_2}{3\ell^2}
+\fft1{\ell^2}\Bigl(
\fft{\mu_1^2(\mu_2-\mu_3)^2+\mu_2^2\mu_3^2}{(\mu_1-\mu_2)(\mu_1-\mu_3)}
\log\mu_1\nonumber\\
&&\qquad
+\fft{\mu_2^2(\mu_1-\mu_3)^2+\mu_1^2\mu_3^2}{(\mu_2-\mu_3)(\mu_2-\mu_3)}
\log\mu_2
+\fft{\mu_3^2(\mu_1-\mu_2)^2+\mu_1^2\mu_2^2}{(\mu_3-\mu_1)(\mu_3-\mu_2)}
\log\mu_3
\Bigr)\Bigr].
\end{eqnarray}
Note the appearance of the logarithmic terms that were not present in the
non-rotating case.

In anticipation of the computation of the boundary stress tensor, we find
that the non-vanishing components of the extrinsic curvature are
\begin{eqnarray}
&&\theta_{00}=\ft12(fg)^{\fft12}\left(-\fft{f'}f+\fft{g'}g-\fft{h'}h\right),
\qquad
\theta_{03}=\fft{f^2w}r\left(\fft2r+\fft{h'}h-\fft{w'}w\right),\nonumber\\
&&\theta_{11}=\theta_{22}=\ft12(fg)^{\fft12}\left(-\fft2r+\fft{f'}f\right),
\qquad
\theta_{33}=\ft12(fg)^{\fft12}\left(-\fft2r-\fft{2f'}f-\fft{h'}h\right),
\end{eqnarray}
so that
\begin{equation}
\theta=\ft12(fg)^{\fft12}\left(-\fft6r+\fft{f'}f-\fft{g'}g\right).
\end{equation}
The trace of the extrinsic curvature is used to compute the Gibbons-Hawking
term $I_{\rm GH}$.  For $I_2$, we also need the intrinsic curvature on
the boundary, ${\cal R}=\fft2{r^2}f(4-hf^3)$.  Adding all contributions
according to (\ref{eq:fcta}), we finally arrive at
\begin{eqnarray}
I_{\rm complete}&=&\fft{\beta\pi}{4G_5}\Bigl[\fft38\ell^2
-\fft{\alpha_2}{2\ell^2}+\fft1{\ell^2}\Bigl(
\fft{\mu_1^2(\mu_2-\mu_3)^2+\mu_2^2\mu_3^2}{(\mu_1-\mu_2)(\mu_1-\mu_3)}
\log\mu_1\nonumber\\
&&\qquad
+\fft{\mu_2^2(\mu_1-\mu_3)^2+\mu_1^2\mu_3^2}{(\mu_2-\mu_3)(\mu_2-\mu_3)}
\log\mu_2
+\fft{\mu_3^2(\mu_1-\mu_2)^2+\mu_1^2\mu_2^2}{(\mu_3-\mu_1)(\mu_3-\mu_2)}
\log\mu_3
\Bigr)\Bigr].\qquad
\end{eqnarray}

\subsection{The boundary stress tensor}

We now proceed to compute the boundary stress tensor and to extract the
ADM energy and angular momentum of this solution.  In fact, the result
is rather simple, and we find
\begin{eqnarray}
T_{00}^{\rm complete}&=&\fft1{8\pi G_5}\fft\ell{r^4}\left(\fft38\ell^2
+\alpha_1+\fft{3\alpha_2}{2\ell^2}+\fft{2\alpha_3}{\ell^4}\right),\nonumber\\
T_{03}^{\rm complete}&=&\fft1{8\pi G_5}\fft\epsilon{\ell r^4}
\left(\alpha_2+\fft{2\alpha_3}{\ell^2}\right).
\end{eqnarray}
In addition, $T_{11}=T_{22}$ and $T_{33}$ are non-vanishing, but do not
contribute to conserved quantities.

Taking into account (\ref{eq:consq}), the conserved ADM energy and angular
momentum are
\begin{eqnarray}
E&=&\fft\pi{4 G_5}\left(\fft38\ell^2+\alpha_1+\fft{3\alpha_2}{2\ell^2}
+\fft{2\alpha_3}{\ell^4}\right),\nonumber\\
J&=&\fft{\epsilon\pi}{8G_5\ell}\left(\alpha_2+\fft{2\alpha_3}{\ell^2}\right).
\label{eq:grmass}
\end{eqnarray}
These expressions agree with those obtained by Gutowski and Reall
\cite{gutreal} using the methods of Ashtekar and Das \cite{ashdas}, provided
one relates the ADM energy $E$ and the Ashtekar and Das mass $M$ through
\begin{equation}
E=M+\fft{3\pi\ell^2}{32G_5}.
\end{equation}
The latter contribution is identified as the Casimir energy, and verifies
the prediction of Gutowski and Reall.


\section{Discussion}

Computing black hole energies using the boundary stress tensor method
is natural in the AdS/CFT context.  What we have shown here is that,
by incorporating a $\phi^2$ counterterm, we are able to derive the expected
ADM energies for the non-rotating $R$-charged black holes, (\ref{eq:goodmass}),
and the rotating BPS solutions, (\ref{eq:grmass}).  In the former case,
this finite counterterm removes a non-linear charge contribution to the
energy, while in the latter case, it modifies but does not remove the
non-linearities.

For the case of the non-rotating black holes, the linear mass relation
(\ref{eq:goodmass}) verifies the result of \cite{Behrndt:1998jd}.  As
this was the basis of the thermodynamic exploration of $R$-charged black
holes in \cite{Cvetic:1999ne}, we have shown that the standard results
follow naturally from the boundary stress tensor prescription, provided
appropriate finite counterterms are incorporated.  The mass of rotating
Einstein-Maxwell AdS$_5$ black holes was also examined using the boundary
stress tensor method in \cite{Klemm:2000gh}.  The result of
\cite{Klemm:2000gh} ought to be generalizable to the STU model after
inclusion of the appropriate $\phi^2$ counterterm.

Of course, as we have indicated, the energy computed in this manner is not
unique, and depends on the nature of finite counterterms used in regulating
the boundary stress tensor. This fact is understood in terms of having to
specify a particular counterterm prescription with which to work with; in a
field theory language, this is simply the scheme dependence of standard
renormalization. Although the energy, as so defined, is ambiguous up to finite
counterterms, physical quantities in the dual field theory must always be
well defined. However, in practice, what is and is not scheme dependent is
often a subtle issue, and separating the two may require care.

In order to deal with this ambiguity, it is natural to impose some additional
symmetry requirements on the regularization procedure.  In the present
case, our desire to expose a linear BPS-like relation between mass and
$R$-charge in the dual CFT has led us to postulate the addition of the
finite $\phi^2$ counterterm in (\ref{eq:fcta}).  In fact, such a counterterm
can be motivated by Hamilton-Jacobi theory, and can be seen as a
necessity for the preservation of supersymmetry in the boundary theory.
Note, also, that for the case of AdS$_4$ with scalars, the $\phi^2$
counterterm is no longer optional, but necessary to render the action
finite.  This connection to the Hamilton-Jacobi approach for matter coupled
gravity systems will be explored in a subsequent publication \cite{wip}.

\section*{Acknowledgments}

This material is based upon work supported by the National Science
Foundation under grant PHY-0313416 and by the US Department of Energy
under grant DE-FG02-95ER40899. JTL wishes to thank A.~Batrachenko, A.~Buchel,
R.~McNees, L.~Pando Zayas and W.Y.~Wen for discussions, and acknowledges
the hospitality of Khuri lab at the Rockefeller University, where part of
this work was completed.

\end{document}